\documentclass[12pt]{iopart}
\usepackage{iopams}

\begin{document}

\title{MUB-like structures and tomographic reconstruction for N-ququart systems}
\author{Juan D\'iaz-Guevara$^{1}$, Isabel Sainz$^{1}$, and Andrei B.
Klimov$^{1,2}$}
\address{$^1$ Departamento de F\'isica, Universidad de Guadalajara, Revoluci\'on 1500,
Guadalajara, Jalisco 44420, Mexico\\
$^2$ Departamento de F\'isica, Center of Quantum Optics and Quantum Information, Center for Optics and Photonics, Universidad de Concepci\'on, Casilla-160C, Concepci\'on, Chile}
\ead{klimov.andrei@gmail.com}
\begin{abstract}
We construct an informationally complete set of mutually unbiased - like bases for N ququarts. These bases are used in an explicit tomographic protocol which performance is analyzed by estimating quadratic errors and compared to other reconstruction schemes.
\end{abstract}

\section{Introduction}
Discrete quantum systems, i.e. systems with a finite number of energy levels
subjected only to Clifford-type physical operations (those that preserve the
generalized Pauli group), play a fundamental role in a wide range of quantum
information protocols as teleportation, quantum key distribution, error
correction codes \cite{tele, QKD, QKD2, ecodes, ecodes1} and in problems
related to state characterization and quantum tomography \cite{charac,tomo}.
In most experiments related to information transmission, only two level
systems (qubits) are involved. However, it has been shown (for a recent
review see \cite{review} and references therein) that quantum information
processing can be optimized by using systems of larger dimension (qudits)
and (effective) non-unitary operations \cite{qudits}. In recent years
special attention has been paid to ququarts (four-level systems), since they
can be efficiently simulated in atomic systems \cite{diamond, diamond1}, in
single photon setups \cite{qqpreparation,permutation,qqlogic} and in nuclear
spins \cite{nmr,permutation2}. Effective ququart systems have been
experimentally applied in quantum computing without entanglement \cite%
{permutation,permutation2}, high-dimensional quantum key distribution \cite%
{qqkd}, and four-dimensional entanglement distribution \cite{ent1,ent2}. In
general, manipulations with a multipartite quantum system of a given
dimension become \textquotedblleft cheaper\textquotedblright , i.e. it is
required less non-local operations, when the dimensionality of its
components increase. For instance, an elementary controlled-ququart gate
costs no more that $8$ CNOT gates, while $104$ CNOT gates are needed for an
implementation of a full four qubit logic \cite{qqlogic}.

One of the fundamental problems in quantum information processing is the
reconstruction of quantum states with a certain fidelity. The lower
intrinsic error generated in tomographic protocols depends both on the type
of the system and on the operations available in a given experimental setup
\cite{error}. A variety of reconstruction schemes for $N$-qubit systems
where proposed and experimentally implemented. The most theoreically
advantageous tomographic method is based on the so-called Mutually Unbiased
Bases (MUB), which can be constructed in multiple ways in a $2^{N}$
dimensional Hilbert space. Most of the states composing the MUBs are highly
entangled and require a large amount of non-local resources for their
generation. Thus, the MUB tomography, while being optimal with respect to
the minimization of statistical errors, turns out to be extremely
\textquotedblleft expensive\textquotedblright , for the experimental
implementation, due to a high entanglement cost of sates used in the
respective projective measurements \cite{wootters, wootters1, ivanovic, mub}. Considering this, the reconstruction of a compound quantum system by using
their (effective) higher dimensional constituents could be an attractive
alternative.

On the other hand, the mathematical construction of MUBs for $N$-qubit (and
any prime power) systems is heavily based on the underlying structure of
finite fields \cite{wootters,mub,band}. That allows to simplify the
inversion of high-rank matrices leading to explicit reconstruction
expressions and a direct quantification of the estimation errors \cite{helstrom}.

The simplest $N$-partite system that can not be treated in this simple way,
i.e. using the standard language of finite fields, is a collection of four
level systems, where $SU(4)$ transformations can be \textit{locally}
implemented. The analysis of a system of $N$ ququarts, and in particular,
the problem of state reconstruction, requires the use of the Galois rings,
whose structure is more involved than the one of the finite fields. This
complexity is particularly reflected in the necessity of \textit{redundant}
measurements for a direct tomographic protocol \cite{vourdas}, \cite{vourdas2}.

In this paper, we construct an informationally complete set of $4^{N}+2^{N}$
mutually unbiased MU-like bases for $N$-ququarts. We also obtain an
explicit state reconstruction expression, which is used for the estimation
of average square error of the Hilbert-Schmidt distance between the real and
the estimated state. We compare the efficiency of our reconstruction scheme
with the corresponding $2N$ qubit MUB tomography.

The paper is organized as follows: in section \ref{nqubittomo} we briefly review $N$-qubit MUB construction and the corresponding tomographic scheme; in section \ref{nququartbases} we construct MU-like bases for single and $N$ ququarts and discuss their properties; in section \ref{nququarttomo} we obtain an explicit reconstruction expression for $N$ ququarts, and analyze its performance. A summary of finite fields and Galois rings can be found in Appendices, for more details see\cite{wang}.

\section{Review for $N$ qubit tomography}\label{nqubittomo}

It is convenient to label the computational basis $\{\left\vert
k_{1}\right\rangle \left\vert k_{2}\right\rangle \cdots \left\vert
k_{N}\right\rangle ,$ $k_{i}=0,1\}$ in the $N$-qubit Hilbert space $\mathcal{%
H}_{2^{N}}=\mathcal{H}_{2}^{\otimes N}$  with elements of the finite field $%
\mathbb{F}_{2^{N}}$, considered as a linear space spanned by a basis $%
\{\theta _{1},...,\theta _{N}\}$, allowing to decompose any $\kappa \in
\mathbb{F}_{2^{N}}$ as,%
\begin{eqnarray}
\kappa =\sum_{i=1}^{N}k_{i}\,\theta _{i}\,,\;k_{i}\in \mathbb{Z}_{2},
\label{ki}
\end{eqnarray}
so that,
\begin{eqnarray}
\left\vert k_{1}\right\rangle \left\vert k_{2}\right\rangle \cdots
\left\vert k_{N}\right\rangle =\left\vert \kappa \right\rangle ,\;\langle
\kappa |\kappa ^{\prime }\rangle =\delta _{\kappa ,\kappa ^{\prime }},
\label{basecompqb}
\end{eqnarray}
see \ref{fields}. In $\mathbb{F}_{2^{N}}$ there always exists the so-called self-dual basis, orthonormal with respect to the trace operation, $\tr(\theta _{i}\,\theta _{j})=\delta _{ij}$,$\;\tr(\kappa )=\kappa +\kappa^{2}+...+\kappa ^{2^{N-1}}\in \mathbb{Z}_{2}$. Thus, in Eq.(\ref{ki}), $k_{i}$
are components of $\kappa$ when it is written in a self-dual basis $k_{i}=\tr(\kappa\theta_{i})$. This allows to establish a correspondence between qubits and elements of the basis in $\mathbb{F}_{2^{N}}$: qubit$_{i}\leftrightarrow
\theta _{i}$, through the trace operation.

The operators, elements of the generalized Pauli group $\mathcal{P}^{N}=\mathcal{P}^{1}\otimes ...\otimes \mathcal{P}^{1}$\ \cite{Schwinger}, \cite{stabilizers}, \cite{discr} acting in $\mathcal{H}_{2}^{\otimes N}$ are defined according to,
\begin{eqnarray}
Z_{\gamma }=\sum_{\kappa\in\mathbb{F}_{2^N} }(-1)^{\mathrm{tr}(\gamma \kappa )}\left\vert
\kappa \right\rangle \langle \kappa |,\quad X_{\delta }=\sum_{\kappa\in\mathbb{F}_{2^N}
}\left\vert \kappa +\delta \right\rangle \langle \kappa |,\;\gamma ,\delta
\in \mathbb{F}_{2^{N}},  \label{ZX}
\end{eqnarray}
satisfying the commutation relations,
\begin{eqnarray}
X_{\delta }Z_{\gamma }=(-1)^{\tr(\delta \gamma )}Z_{\gamma
}X_{\delta },  \label{dualqb}
\end{eqnarray}
and are related through the finite Fourier transform operation, $X_{\delta
}=F_{2^{N}}^{-1}Z_{\delta }F_{2^{N}},$
\begin{eqnarray}
F_{2^{N}}=2^{-N/2}\sum_{\alpha ,\beta \in \mathbb{F}_{2^{N}}}(-1)^{\tr(\alpha\beta )}|\alpha\rangle\langle \beta|. 
\end{eqnarray}
The operators (\ref{ZX}) are factorized into a direct product of single qubit operators,
\begin{eqnarray}
Z_{\gamma} &=\sigma _{z}^{g_{1}}\otimes \cdots \otimes \sigma
_{z}^{g_N},\quad g_{i}=\tr(\gamma \theta_{i}),  \nonumber \\
X_{\delta} &=\sigma _{x}^{d_{1}}\otimes \cdots \otimes \sigma
_{x}^{d_N},\quad d_{i}=\tr(\delta \theta_{i}),
\end{eqnarray}
as well as Fourier operators%
\begin{eqnarray*}
F_{2^{N}}=F_{2}\otimes \cdots \otimes F_{2},\quad F_{2}=2^{-1/2}\sum_{\ell,\ell'=0}^{1}(-1)^{\ell \,\ell'}\,|\ell \rangle \langle \ell '|\,
\end{eqnarray*}
where $\sigma _{x,y,z}$ are Pauli operators.
The set of $2^{2N}$ monomials $\{Z_{\gamma }X_{\delta }$, $\gamma ,\delta
\in \mathbb{F}_{2^{N}}\}$ form an operational basis in $\mathcal{H}_{2}^{\otimes N}$ and can be separated into $2^{N}+1$ subsets of $2^{N}$
commuting monomials,
\begin{eqnarray}
\{Z_{\gamma }X_{\lambda \gamma },\;[Z_{\gamma }X_{\lambda \gamma },Z_{\gamma
^{\prime }}X_{\lambda \gamma ^{\prime }}]=0,\;\lambda ,\gamma ,\gamma
^{\prime }\in \mathbb{F}_{2^{N}}\}\cup \{X_{\gamma },\;\gamma \in \mathbb{F}_{2^{N}}\}.  \label{US}
\end{eqnarray}
The eigenstates of the commuting sets (\ref{US}),
\begin{eqnarray}
Z_{\gamma }X_{\lambda \gamma }|\psi _{\kappa }^{\lambda
}\rangle &=(-1)^{\tr(\gamma \kappa )}| \psi_{\kappa
}^{\lambda }\rangle ,\quad \mathrm{ fixed }~\lambda ,  \label{MUB2Ma} \\
X_{\gamma }|\tilde{\kappa}\rangle &=(-1)^{\tr(\gamma \kappa )}| \tilde{\kappa}\rangle ,\quad |\tilde{\kappa}\rangle =F_{2^{N}}^{-1}|\kappa \rangle ,
\label{MUB2Mb}
\end{eqnarray}
are mutually unbiased \cite{wootters, wootters1, ivanovic, mub, band},
\begin{eqnarray}
|\langle \psi _{\kappa'}^{\lambda'}|\psi _{\kappa}^{\lambda}\rangle |^{2} &=\delta _{\lambda \lambda'}\delta_{\kappa \kappa'}+\frac{1-\delta _{\lambda \lambda'}}{2^{N}},  \label{MUB2a} \\
\langle \tilde{\kappa}'\left\vert \tilde{\kappa}\right\rangle &=
\delta_{\kappa ,\kappa'},\quad |\langle \tilde{\kappa}^{\prime
}|\psi _{\kappa }^{\lambda }\rangle |^{2}=\frac{1}{2^{N}}.  \label{MUB2b}
\end{eqnarray}
Explicitly, the eigenstates of a set $\{Z_{\gamma }X_{\lambda \gamma }$,
fixed $\lambda \}$ are obtained from elements of the computational basis $%
|\kappa\rangle $ through the following unitary transformation,
\begin{eqnarray}
|\psi _{\kappa }^{\lambda }\rangle &= V_{\lambda }|
\kappa \rangle ,\quad V_{\lambda }=\frac{1}{2^{N}}\sum_{\alpha ,\beta
,\gamma \in \mathbb{F}_{2^{N}}}c_{\gamma ,\lambda }(-1)^{\tr\left(
\gamma (\alpha -\beta )\right) }| \alpha\rangle\langle\beta| ,  \label{2mub} \\
V_{\lambda }Z_{\gamma }V_{\lambda }^{\dag } &= c_{\gamma ,\lambda}Z_{\gamma
}X_{\lambda \gamma},
\end{eqnarray}
where the phases $c_{\gamma,\lambda}$, $|c_{\gamma,\lambda }|=1,$ satisfy
the functional equation,
\begin{eqnarray}
c_{\alpha +\gamma ,\lambda }=c_{\alpha,\lambda}c_{\gamma,\lambda }(-1)^{\tr(\alpha \gamma \lambda )}. \label{phaseqb}
\end{eqnarray}
As discussed in \ref{galois}, a set of solutions of (\ref{phaseqb}) can be easily obtained considering the indices of $c_{\gamma ,\lambda}$ as elements of the ring $GR(4,N)$, $\gamma ,\lambda \in\mathcal{T}_{2}=GR(4,N)/(2)\subset GR(4,N)$ which are in one-to-one correspondence with $\mathbb{F}_{2^{N}}$.

The density matrix has a simple expansion on the projectors on the MUBs
\begin{eqnarray}
\rho =\sum_{\lambda ,\kappa \in \mathbb{F}_{2^{2N}}}p_{\kappa }^{\lambda
}|\psi _{\kappa }^{\lambda }\rangle \langle \psi _{\kappa }^{\lambda
}|+\sum_{\kappa \in \mathbb{F}_{2^{2N}}}\tilde{p}_{\kappa }|\tilde{\kappa}\rangle \langle \tilde{\kappa}|-\mathbb{I},  \label{qbformula}
\end{eqnarray}
where the measured probabilities $p_{\kappa }^{\lambda }=\langle \psi
_{\kappa }^{\lambda }|\rho |\psi _{\kappa }^{\lambda }\rangle $ and $\tilde{p}_{\kappa }=\langle \tilde{\kappa}|\rho |\tilde{\kappa}\rangle $ satisfy the normalization conditions,
\begin{eqnarray}
\sum_{\kappa \in \mathbb{F}_{2^{2N}}}p_{\kappa }^{\lambda }=\sum_{\kappa \in
\mathbb{F}_{2^{2N}}}\tilde{p}_{\kappa }=1.  \label{NC}
\end{eqnarray}
The above reconstruction scheme is optimal in the sense of minimisation of
the statistical error associated with the $2^{N}+1$ measurement setups.

Unfortunately, among all MUBs there are only $3$ completely factorized
bases, eigenstates of $\{|\psi _{\kappa }^{\lambda =0}\rangle \equiv
|\kappa\rangle \},\|\tilde{\kappa}\rangle
\}$ and $\{|\psi _{\kappa }^{\lambda =1}\rangle \}$ which are eigenstates of
the sets $\{Z_{\gamma }\}$, $\{X_{\gamma }\}$ and $\{Z_{\gamma }X_{\gamma
}\}$ correspondingly. All the other MUBs of the type (\ref{2mub}) are
entangled, which makes the experimental realization of MUB tomography very
\textquotedblleft expensive\textquotedblright\ \cite{klimov}, \cite{klimov1}.

\section{N ququart MU-like bases}\label{nququartbases}

\subsection{Single ququart}

Let us consider a single particle with four energy levels (ququart).
Although the dimension of the Hilbert space for a quaquart is the same as
for 2 qubits $\mathcal{H}_{4}=\mathcal{H}_{2}^{\otimes 2}$, there is a
crucial difference in the type of operations available in both systems:
non-local transformations in two-qubit systems correspond to local $SU(4)$
operations in ququart systems. Here we focus only on discrete
transformations generated by unitary operators
\begin{eqnarray}
Z=\sum_{k=0}^{3}i^{k}|k\rangle \langle k|,\quad X=\sum_{k=0}^{3}|k+1\rangle
\langle k|,  \label{ZX4}
\end{eqnarray}
where the algebraic operations are $\mathrm{mod}~4$. The operators (\ref{ZX4}) are cyclic, $Z^{4}=X^{4}=\mathbb{I}$, and satisfy the commutation relation,
\begin{eqnarray}
Z^{a}X^{b}=i^{ab}X^{b}Z^{a},\quad a,b\in \mathbb{Z}_{4}.  \label{comm1}
\end{eqnarray}
The full set of $16$ monomials $\{Z^{a}X^{b},a,b\in \mathbb{Z}_{4}\}$ form an
operational basis in $\mathcal{H}_{4}$ \cite{Schwinger}, and can be grouped
into $6$ commuting sets, which in contrast to qubit systems are not disjoint.
There are two types of commuting sets:

i) four sets of the form,
\begin{eqnarray}
\{{Z^{a}X^{la},}a\in \mathbb{Z}_{4}\},\quad l=0,1,2,3;  \label{UR1}
\end{eqnarray}

ii) two sets of the form,
\begin{eqnarray}
\{Z^{mb}X^{b},b\in \mathbb{Z}_{4}\},\quad m=0,2;  \label{UR2}
\end{eqnarray}
see table \ref{monomials1qq}. The sets (\ref{UR1}) and (\ref{UR2}) are not
equivalent since $m=0,2$ are zero divisors, i.e. have no multiplicative
inverse in $\mathbb{Z}_{4}$. One can observe that only $3$ mutually disjoint
sets can be chosen out of $6$.
\begin{table}
\caption{Table of commuting monomials for one ququart. Each row corresponds
to a commuting set. The first four rows corresponds to the sets ${Z^{a}X^{la}}$, with $l\in \mathbb{Z}_{4}$ and the last two rows correspond to the sets $Z^{mb}X^{b}$, with $m\in (2)=\{0,2\}$.}
\label{commutingsets}
\begin{center}
\begin{tabular}{|c|c|c||c|c|}
\hline
$Z$ & $Z^2$ & $Z^3$ & $l=0$ & $Z^aX^{(l=0)a}$ \\ \hline
$ZX^2$ & $Z^2$ & $Z^3X^2$ & $l=2$ & $Z^aX^{(l=2)a}$ \\ \hline
$ZX$ & $Z^2X^2$ & $Z^3X^3$ & $l=1$ & $Z^aX^{(l=1)a}$ \\ \hline
$ZX^3$ & $Z^2X^2$ & $Z^3X$ & $l=3$ & $Z^aX^{(l=3)a}$ \\ \hline
$X$ & $X^2$ & $X^3$ & $m=0$ & $Z^{(m=0)b}X^{b}$ \\ \hline
$Z^2X$ & $X^2$ & $Z^2X^3$ & $m=2$ & $Z^{(m=2)b}X^{b} $ \\ \hline
\end{tabular}
\end{center}

\label{monomials1qq}
\end{table}

The bases corresponding to commuting sets $\{Z^{a}X^{la}\}$ are labelled as $\{|\psi_{k}^{l}\rangle$, $k\in \mathbb{Z}_{4}\}$, being the computational basis $\{|k\rangle =|\psi_{k}^{l=0}\rangle ,k\in \mathbb{Z}_{4}\}$; the eigenstates of the sets $\{Z^{mb}X^{b}\}$ are $\{|\tilde{\psi}_{k}^{m}\rangle, k\in \mathbb{Z}_{4}\}$. According to the general approach, the bases
corresponding to disjoint sets are unbiased \cite{band}. The overlap relations between eigenstates of sets that share operators are more involved, but possess certain symmetries, that can be used for constructing tomographic protocols. The explicit form of all $6$ bases is given in \ref{apenC}.

The overlap relations between the elements of the bases can be represented
in a compact form by introducing the \textit{bar map} \cite{wang}: $~\bar{}:%
\mathbb{Z}_{4}\rightarrow \mathbb{Z}_{2}$ which is defined as $\overline{a}%
=a~ \mathrm{mod}~ 2$, $\forall a\in \mathbb{Z}_{4}$.

\begin{enumerate}
\item The eigenbases of $\{Z^{a}X^{la}\}$ are unbiased to the eigenstates of
$\{Z^{mb}X^{b}\}$ for every $l=0,\ldots,3$, and $m=0,2$,
\begin{eqnarray*}
|\langle \psi_{k}^{l}|\tilde{\psi}_{k'}^{m}\rangle |^{2}=\frac{1}{4},~~\forall k,k'\in \mathbb{Z}_{4}.
\end{eqnarray*}
\item The eigenstates of $\{Z^{a}X^{la}\}$ and $\{Z^{a}X^{l'a}\}$
are mutually unbiased if and only if $\overline{l}\neq \overline{l'}
$; if $\overline{l}=\overline{l'}$, the overlap is either $0$ or $%
1/2$,
\begin{eqnarray*}
|\langle \psi _{k}^{l}|\psi _{k'}^{l'}\rangle |^{2}=\delta
_{kk'}\delta _{ll'}+(1-\delta _{ll'})\left(\frac{\delta _{\overline{k},\overline{k'}}\delta _{\overline{l},\overline{l'}}}{2}+\frac{1-\delta _{\overline{l},\overline{l'}}}{4}\right) ,~~k,k'\in \mathbb{Z}_{4}.
\end{eqnarray*}
\item The eigenbases of $\{X^{b}\}$ and $\{Z^{2b}X^{b}\}$ are not unbiased
and satisfy the relation,
\begin{eqnarray*}
|\langle \tilde{\psi}_{k}^{0}|\tilde{\psi}_{k'}^{2}\rangle |^{2}=\frac{1}{2}\delta _{\bar{k},\bar{k'}},~~k,k'\in \mathbb{Z}_{4}.
\end{eqnarray*}
Thus, there are only $3$ MUBs among $6$ bases, according to the number of
disjoint sets.
\end{enumerate}

\subsection{N ququarts}

In case of $N$ ququarts it is convenient to label both states and operators
acting in the Hilbert space $\mathcal{H}_{4^{N}}=\mathcal{H}_{4}^{\otimes N}$, with elements of the Galois Ring $GR(4,N)$, that is considered as a linear
space spanned by a basis $\{\theta _{i},i=1,..,N\}$,
\begin{eqnarray}
\alpha =a_{1}\theta _{1}+a_{2}\theta _{2}+...+a_{N}\theta _{N},\quad \alpha
\in GR(4,N),\quad a_{i}\in \mathbb{Z}_{4}.  \label{basisexpantion}
\end{eqnarray}
The components in the above decomposition are related to the ring element
through the trace operation $\mathrm{T}_{4}:GR(4,N)\rightarrow \mathbb{Z}_{4}$,
\begin{eqnarray}
a_{i}=\mathrm{T}_{4}(\alpha \theta _{i}^{\ast }),  \label{t4}
\end{eqnarray}
being $\{\theta_{i}^*\}$ the basis dual to $\{\theta _{i}\}$, i.e. $\mathrm{T}_{4}(\theta _{i}\theta _{j}^*)=\delta _{ij}$, see \ref
{galois1}, and each element of the basis can be put in correspondence with
a ququart. Elements of the ring are separated into units, having a
multiplicative inverse, and zero divisors.

The computational basis in $\mathcal{H}_{4^{N}}$ is formed by the states
\begin{eqnarray}
|k_{1}\rangle \otimes ...\otimes |k_{N}\rangle =|\kappa \rangle ,\quad
k_{i}=\mathrm{T}_{4}(\kappa \theta _{i}^{\ast })\in \mathbb{Z}_{4},\quad \kappa \in
GR(4,N).  \label{logic basis}
\end{eqnarray}
The vectors (\ref{logic basis}) are eigenstates of the set of $4^{N}$ commuting operators,
\begin{eqnarray}
Z_{\gamma }=\sum_{\kappa \in GR(4,N)}i^{\mathrm{T}_{4}(\gamma \kappa )}|\kappa\rangle \langle \kappa |,\quad Z_{\gamma }=Z^{g_{1}}\otimes ...\otimes
Z^{g_{N}},  \label{ZR}
\end{eqnarray}
where $\gamma=g_1\theta^*_1+\cdots+g_N\theta^*_N \in GR(4,N)$,  $g_{j}=\mathrm{T}_{4}(\gamma \theta _{j})\in \mathbb{Z}_{4}$, and a single ququart operator $Z$ is defined by (\ref{ZX4}). From now on, all the sums run over $GR(4,N)$, unless otherwise specified. The shift operators,
\begin{eqnarray}
X_{\delta }=X^{d_{1}}\otimes ...\otimes X^{d_{N}},\quad X_{\delta
}=\sum_{\kappa }|\kappa +\delta \rangle \langle \kappa |,  \label{XR}
\end{eqnarray}
where $\delta =d_1\theta_1+\cdots+d_N\theta_N\in GR(4,N)$, $d_{j}=\mathrm{T}_{4}(\delta \theta _{j}^*)\in\mathbb{Z}_{4}$, and the operator $X$ defined in (\ref{ZX4}), form a complementary set to (\ref{ZR}),
\begin{eqnarray}
X_{\delta }Z_{\gamma }=i^{T_{4}(\delta \gamma )}Z_{\gamma }X_{\delta }.
\label{comm}
\end{eqnarray}
Observe that if $\delta=\gamma$, the powers $\{d_{j},j=1,...,N\}$ in (\ref{XR}) are related to $\{g_{j},j=1,...,N\}$ in (\ref{ZR}) according to $d_{j}=\sum_{i=1}^{N}g_{i}\mathrm{T}_{4}(\theta _{i}^*\theta _{j}^*)$. Thus, if $\delta=\gamma$, $d_{j}=g_{j}$ only in the case of decomposition on the self-dual basis. This, in particular, entails that the Fourier operator
\begin{eqnarray}
F_{4^{N}}=\frac{1}{2^{N}}\sum_{\alpha ,\beta }i^{\mathrm{T}_{4}(\alpha \beta
)}|\alpha \rangle \langle \beta |,\quad X_{\delta }=F_{4^{N}}^{-1}Z_{\delta
}F_{4^{N}},  \label{fourier}
\end{eqnarray}
is factorized only in a self-dual basis. A self-dual basis exists in $GR(4,N)
$ only for odd $N$ values \cite{bagio}.

The monomials $Z_{\gamma }X_{\delta }$ form an operational basis in $\mathcal{H}_{4^{N}}$ and can be separated into two types of commuting sets,
\begin{eqnarray}
\{Z_{\gamma }X_{\lambda \gamma }|~\gamma ,\lambda \in GR(4,N)\},  \label{CS1}
\\
\{Z_{\mu \delta }X_{\delta }|~\delta \in GR(4,N),~\mu \in (2)\},  \label{CS2}
\end{eqnarray}
labelled with the indices $\lambda \in GR(4,N)$ and zero-divisors $\mu $
contained in the principal ideal $(2)$ of $GR(4,N)$. Since there are $2^{N}$
zero divisors in $GR(4,N)$, there exist in total $4^{N}+2^{N}$ commuting
sets.

Not all the sets (\ref{CS1})-(\ref{CS2}) are disjoint. It immediately
follows from (\ref{comm}) that,

i) the sets (\ref{CS1}) are disjoint only if $\overline{\lambda }\neq
\overline{\lambda'}$, where the bar map, $GR(4,N)\rightarrow
\mathbb{F}_{2^{N}}$, is defined as $\overline{\lambda }=\lambda~ \mathrm{mod}~2$; Thus, the sets $\{Z_{\gamma }X_{\lambda \gamma }\}$ and $\{Z_{\gamma
}X_{\lambda'\gamma }\}$ with $\overline{\lambda}=\overline{\lambda'}$ share the elements $\{Z_{\gamma }X_{\overline{\lambda }\gamma }|~\gamma \in (2)\}$;

ii) any two sets $\{Z_{\mu \delta }X_{\delta }\}$ and $\{Z_{\mu ^{\prime
}\delta }X_{\delta }\}$ are not disjoint since $\overline{\mu }=0$ if $\mu
\in (2)$, and share the elements $\{X_{\delta }|~\delta \in (2)\};$

iii) the sets (\ref{CS1}) and (\ref{CS2}) are mutually disjoint.

Thus, there are $2^{N}+1$ groups each containing $2^{N}$ non disjoint sets
of commuting monomials, where the sets from different groups are \ mutually
disjoint. In a sense, this is a generalization of the $N$-qubit case, where
each group contains only one commuting set. In Table II the structure of
commuting sets (\ref{CS1})-(\ref{CS2}) for two ququarts is presented.

The above structure can be nicely represented in a non-Euclidian finite
plane, where the axis ($\gamma ,\delta $) are labeled by the indexes of $%
Z_{\gamma }$ and $X_{\delta }$ operators. In the geometrical representation
the commuting monomials (\ref{CS1})-(\ref{CS2}) are put in correspondence
with two inequivalent bundles of rays $\delta =\lambda \gamma $, $\lambda
\in GR(4,N)$ and $\lambda \delta =\gamma $ , $\lambda \in (2)$,
respectively. However, in contrast to prime power dimensions \cite{wootters1}, the rays with the same $\overline{\lambda }$ are intersected in sublines $\delta =\overline{\lambda }\gamma $, when $\lambda \in GR(4,N)$ and $\gamma=0$ if $\lambda \in (2)$ \cite{vourdas,vourdas2}.

\begin{table}
\caption{Table of the sets of commuting operators for two ququarts. Each
entry corresponds to one set. In every row, the sets are not disjoint. First
four rows correspond to the sets $\{Z_{\protect\gamma }X_{\protect\lambda
\protect\gamma }\}$, the last row corresponds the sets $\{Z_{\protect\mu
\protect\delta }X_{\protect\delta }\}$. The right column contains the
operators shared between the sets of each row. $\protect\xi $ is a root of
the irreducible polynomial on $\mathbb{Z}_{4}:$ $x^{2}+x+1=0.$}
\label{monomials2qq}
\begin{center}
\begin{tabular}{|c|c|c|c||c|c|}
\hline
$Z_{\gamma }$ & $Z_{\gamma }X_{2\gamma }$ & $Z_{\gamma }X_{2\xi \gamma }$ & $%
Z_{\gamma }X_{2\xi ^{2}\gamma }$ & $\overline{\lambda }=0$ & $Z_{2},Z_{2\xi
},Z_{2\xi ^{2}}$ \\ \hline
$Z_{\gamma }X_{\gamma }$ & $Z_{\gamma }X_{3\gamma }$ & $Z_{\gamma
}X_{(1+2\xi )\gamma }$ & $Z_{\gamma }X_{(1+2\xi ^{2})\gamma }$ & $\overline{%
\lambda }=1$ & $Z_{2}X_{2},Z_{2\xi }X_{2\xi },Z_{2\xi ^{2}}X_{2\xi ^{2}}$ \\
\hline
$Z_{\gamma }X_{\xi \gamma }$ & $Z_{\gamma }X_{(\xi +2)\gamma }$ & $Z_{\gamma
}X_{3\xi \gamma }$ & $Z_{\gamma }X_{(\xi +2\xi ^{2})\gamma }$ & $\overline{%
\lambda }=\xi $ & $Z_{2}X_{2\xi },Z_{2\xi }X_{2\xi ^{2}},Z_{2\xi ^{2}}X_{2}$
\\ \hline
$Z_{\gamma }X_{\xi ^{2}\gamma }$ & $Z_{\gamma }X_{(\xi ^{2}+2)\gamma }$ & $%
Z_{\gamma }X_{(\xi ^{2}+2\xi )\gamma }$ & $Z_{\gamma }X_{3\xi ^{2}\gamma }$
& $\overline{\lambda }=\xi ^{2}$ & $Z_{2}X_{2\xi ^{2}},Z_{2\xi
}X_{2},Z_{2\xi ^{2}}X_{2\xi }$ \\ \hline\hline
$X_{\delta }$ & $Z_{2\delta }X_{\delta }$ & $Z_{2\xi \delta }X_{\delta }$ & $%
Z_{2\xi ^{2}\delta }X_{\delta }$ & $\overline{\mu }=0$ & $X_{2},X_{2\xi
},X_{2\xi ^{2}}$ \\ \hline
\end{tabular}
\end{center}
\end{table}
In practice, it is frequently convenient to use \emph{2-adic notation} (see
\ref{galois2}) for labelling the operators $Z_{\alpha }$ ($%
X_{\alpha })$: $\alpha \in GR(4,N)$ is uniquely represented in terms of the
Teichmuller set, $\mathcal{T}_{2}=\{0,\xi ,\xi ^{2},\ldots ,\xi
^{2^{N}-2},\xi ^{2^{N}-1}=1\}$, as $\alpha =a+2b$, for $a,b\in \mathcal{T}_{2}$.

\subsection{MU-like structure}

The eigenstates of commuting monomials form peculiar bases, that resemble
the properties of MUBs (\ref{MUB2Ma})-(\ref{MUB2b}) in the $N$-qubit case.
The eigenstates of the commuting sets $\{Z_{\gamma }X_{\lambda \gamma }\}$, $\lambda \in GR(4,N)$ are obtained as unitary transformations of the
computational basis (\ref{logic basis}),
\begin{eqnarray}
|\psi _{\kappa }^{\lambda }\rangle =V_{\lambda }|\kappa \rangle ,\quad
V_{\lambda }=\frac{1}{4^{N}}\sum_{\alpha ,\alpha ^{\prime },\beta }c_{\beta
,\lambda }i^{\mathrm{T}_{4}(\beta (\alpha -\alpha ^{\prime }))}|\alpha \rangle
\langle \alpha ^{\prime }|,  \label{4zx}
\end{eqnarray}
where $c_{\gamma ,\lambda }$, $|c_{\gamma ,\lambda }|=1$ satisfy the
following functional equation,
\begin{eqnarray}
c_{\alpha +\gamma ,\lambda }c_{\gamma ,\lambda }^*=c_{\alpha ,\lambda
}i^{3\mathrm{T}_{4}(\alpha \gamma \lambda )}.  \label{fasetomo}
\end{eqnarray}
According to the general method (see \ref{galois4}), a solution of
(\ref{fasetomo}) can be found considering the indices of $c_{\kappa ,\lambda
}$ as elements of $\mathcal{T}_{3}=GR(8,N)/(4)$: $c_{\gamma ,\lambda
}=\omega ^{7\mathrm{T}_{8}(\lambda \gamma ^{2})}$, where $\omega =(1+i)/\sqrt{2}$ is the eighth root of unity and $\mathrm{T}_{8}$ is the trace operation defined in $GR(8,N)$ over $\mathbb{Z}_{8}$, see \ref{galois1}. From now on, we will use indistinctly $|\kappa \rangle $ or $|\psi _{\kappa }^{0}\rangle $ for the computational basis.

The eigenstates $\{|\tilde{\kappa}\rangle $, $\kappa \in GR(4,N)\}$ of the
set $\{X_{\delta }\}$ are obtained as the Fourier transform of the
computational basis (\ref{logic basis}),
\begin{eqnarray}
|\tilde{\kappa}\rangle =F_{4^{N}}^{-1}|\kappa \rangle.  \label{kF}
\end{eqnarray}
The eigenstates $\{|\tilde{\psi}_{\kappa }^{\mu }\rangle \}$ of the sets $\{Z_{\mu \delta }X_{\delta },\mu \in (2)\}$ are unitary equivalent to $\{|\tilde{\kappa}\rangle =|\tilde{\psi}_{\kappa }^{0}\rangle \}$ according to,
\begin{eqnarray}
|\tilde{\psi}_{\kappa }^{\mu }\rangle =\tilde{V}_{\mu }^{\dag }|\tilde{\kappa}\rangle ,\quad \tilde{V}_{\lambda }=F_{4^{N}}^{-1}V_{\lambda}F_{4^{N}}.\label{4xz}
\end{eqnarray}
Thus, the spectral decomposition of monomials $Z_{\gamma }X_{\lambda \gamma
} $ and $Z_{\mu \delta }X_{\delta }$ has the form,
\begin{eqnarray}
Z_{\gamma }X_{\lambda \gamma }=c_{\gamma ,\lambda }^*\sum_{\eta
}i^{\mathrm{T}_{4}(\gamma \eta )}|\psi _{\eta }^{\lambda }\rangle \langle \psi _{\eta}^{\lambda }|,\quad Z_{\mu \delta }X_{\delta }=c_{\delta ,\mu }^*\sum_{\eta}i^{\mathrm{T}_{4}(\delta \eta)}|\tilde{\psi}_{\eta }^{\mu}\rangle
\langle \tilde{\psi}_{\eta }^{\mu }|.  \label{spec}
\end{eqnarray}

The overlap relations between the bases (\ref{4zx}) and (\ref{4xz}) can be
easily found:

\begin{itemize}
\item The bases $\{|\psi_{\kappa}^{\lambda}\rangle \}$ and $\{|\psi
_{\kappa }^{\lambda'}\rangle \}$, where $\lambda ,\lambda'\in GR(4,N),$ are unbiased if $\overline{\lambda }\neq \overline{\lambda'}$. If $\overline{\lambda }=\overline{\lambda'}$, the squared overlap is either $0$ or $2^{-N}$ depending on whether the bar maps of $\kappa $ and $\eta $ are the same or not,
\begin{eqnarray}
|\langle \psi _{\kappa }^{\lambda }|\psi _{\eta }^{\lambda'}\rangle |^{2}=\delta _{\kappa ,\eta }\delta _{\lambda ,\lambda'}+(1-\delta _{\lambda ,\lambda'})\left( \frac{\delta _{\overline{\kappa },\overline{\eta }}\delta _{\overline{\lambda},\overline{\lambda'}}}{2^{N}}+\frac{1-\delta _{\overline{\lambda },\overline{\lambda'}}}{4^{N}}\right).  \label{4ll}
\end{eqnarray}

\item The bases $\{|\psi _{\kappa }^{\lambda }\rangle $, $\lambda \in GR(4,N)\}$ and $\{|\tilde{\psi}_{\kappa }^{\mu }\rangle $, $\mu \in (2)\}$ are unbiased,
\begin{eqnarray}
|\langle \psi _{\kappa }^{\lambda }|\tilde{\psi}_{\eta }^{\mu }\rangle |^{2}=\frac{1}{4^{N}}.  \label{4lm}
\end{eqnarray}

\item The bases $\{|\psi _{\kappa }^{\mu }\rangle \}$ and $\{|\psi _{\kappa
}^{\mu'}\rangle \}$, where $\mu ,\mu'\in (2)$, are not unbiased. The square overlap is either $0$ or $2^{-N}$ depending on whether the bar maps of $\kappa $ and $\eta $ are the same or not,
\begin{eqnarray}
|\langle \tilde{\psi}_{\kappa }^{\mu }|\tilde{\psi}_{\eta }^{\mu'}\rangle |^{2}=\delta_{\kappa,\eta }\delta _{\mu ,\mu'}+(1-\delta _{\mu,\mu'})\frac{\delta _{\bar{\kappa},\bar{\eta}}}{2^{N}}.  \label{4mm}
\end{eqnarray}

\item There are $2^{N}+1$ MUBs among $4^{N}+2^{N}$bases, with $%
2^{N}(2^{N}+1) $ possibilities of choosing them.
\end{itemize}

In other words, the eigenstates of disjoint sets are mutually unbiased,
while sets sharing elements have either orthogonal eigenstates or their
overlap is a constant (up to a phase). As a consequence, the projectors on
the bases (\ref{4zx}) and (\ref{4xz}) satisfy the following redundancy
conditions
\begin{eqnarray}
\sum_{\gamma \in (2)}|\psi _{\bar{\kappa}+\gamma }^{\bar{\lambda}+\delta
}\rangle \langle \psi _{\bar{\kappa}+\gamma }^{\bar{\lambda}+\delta }|
&=\sum_{\gamma \in (2)}|\psi _{\bar{\kappa}+\gamma }^{\bar{\lambda}}\rangle
\langle \psi _{\bar{\kappa}+\gamma }^{\bar{\lambda}}|,  \label{RC1} \\
\sum_{\gamma \in (2)}|\tilde{\psi}_{\bar{\kappa}+\gamma }^{\mu }\rangle
\langle \psi _{\bar{\kappa}+\gamma }^{\mu }| &=\sum_{\gamma \in (2)}|\tilde{
\psi}_{\bar{\kappa}+\gamma }^{0}\rangle \langle \tilde{\psi}_{\bar{\kappa}+\gamma }^{0}|,  \label{RC2}
\end{eqnarray}
where $\delta ,\mu \in (2)$, which also differentiate them from the mutually
unbiased bases in the $N$-qubit case. In composite dimensions a construction
similar to (\ref{4ll})-(\ref{4mm}) is  called weak MUBs \cite{vourdas}, \cite{vourdas2}.

The factorization structure of the bases (\ref{4zx}) and (\ref{4xz}) is
similar to the qubits case. There are four factorized bases of the type (\ref{4zx}), and two factorized bases of the type (\ref{4xz}). All the other
bases are non-factorized. This can be seen straightforward to see for an odd
number of qubits, when there exists a self-dual basis $\{\theta _{i}\}$, $i=1,\ldots ,N$ in $GF(4,N)$, so that the monomials (\ref{CS1}) for $\lambda
=l=0,1,2,3$ \ and (\ref{CS2}) for $\mu =m=0,2$ are factorized according to,
\begin{eqnarray*}
Z_{\gamma }X_{l\gamma } &=Z^{g_{1}}X^{lg_{1}}\otimes \dots \otimes
Z^{g_{N}}X^{lg_{N}},\quad \gamma =\sum g_{i}\theta _{i},\quad g_{i}\in
\mathbb{Z}_{4}, \\
Z_{m\delta }X_{\delta } &=Z^{md_{1}}X^{d_{1}}\otimes \dots \otimes
Z^{md_{N}}X^{d_{N},\quad }\delta =\sum d_{i}\theta _{i},\quad d_{i}\in
\mathbb{Z}_{4}.
\end{eqnarray*}
The monomials in each of the sets above commute by particle, which
guarantees that their eigenstates, $\{|\psi_{\kappa }^{\lambda }\rangle
,\lambda =0,1,2,3\}$ $\{|\tilde{\psi}_{\kappa }^{\mu }\rangle $, $\mu =0,2\}$
are factorized.

In case of even number of ququarts the elements $GR(4,N)$, that label the
factorized bases, depend on the choice of the basis in the ring. For
instance, for two ququarts, the sets (\ref{CS1}) with $\lambda =0,2,\xi
+3\xi ^{2},3\xi +\xi ^{2}$, where $\xi $ is a root of the irreducible
polynomial $x^{2}+x+1$, are factorized when the particles are associated to
the elements of the basis $\{\theta _{1}=\xi ,\theta _{2}=\xi ^{2}\}$. It is
worth noting that in this case the rotation operator (\ref{4zx}) is proportional to the $CNOT_{4}$ operator,
\begin{eqnarray*}
V_{\lambda }\sim CNOT_{4}^{l_{1}+l_{2}},\;\lambda =l_{1}\theta
_{1}+l_{2}\theta_{2},
\end{eqnarray*}
where,
\begin{eqnarray}
CNOT_{4}=\sum_{k_{1}=0}^{3}|\tilde{k}_{1}\rangle_{1\,1}\langle \tilde{k}_{1}| \otimes X_{2}^{k_{1}}.\label{4CNOT}
\end{eqnarray}

\section{MU-like ququart tomography}\label{nququarttomo}

\subsection{General reconstruction formula}

In the $N$-qubit case $2^{N}-1$ independent measurements in each of $2^{N}+1$
(unbiased) bases determine $2^{2N}-1$ entries of the density matrix, in such
a way that every measured probability defines one matrix element according
to the reconstruction equation (\ref{qbformula}). The mutually unbiased-like
structure (\ref{4ll})-(\ref{4mm}) allows to construct a tomographic scheme
for $N$-ququarts. However, $4^{N}-1$ measured probabilities in each of $4^{N}+2^{N}$ bases lead to an informationally overcomplete reconstruction of
$4^{2N}-1$ independent parameters, that determine the $N$-ququart density
matrix.

The density matrix $\rho $ is expanded in the monomial operational basis (\ref{CS1})-(\ref{CS2}) as follows,
\begin{eqnarray}
\rho &=\sum_{\mu \in (2)}\sum_{\kappa \in GR(4,N)/\{0\}}\tilde{A}_{\kappa
}^{\mu }Z_{\mu \kappa }X_{\kappa }+\sum_{\lambda }\sum_{\kappa \in
GR(4,N)/\{0\}}A_{\kappa }^{\lambda }Z_{\kappa }X_{\lambda \kappa }+\frac{\mathbb{I}}{4^{N}}  \nonumber \\
&-\frac{2^{N}-1}{2^{N}}\left( \sum_{\mu \in (2)}\sum_{\kappa \in (2)/\{0\}}\tilde{A}_{\kappa }^{\mu }Z_{\mu \kappa }X_{\kappa }+\sum_{\lambda}\sum_{\kappa \in (2)/\{0\}}A_{\kappa }^{\lambda }Z_{\kappa }X_{\lambda
\kappa }\right) .  \label{redun}
\end{eqnarray}
The last term in (\ref{redun}) takes into account the repetitions present in
the first two terms, i.e. every monomial appears only one time in the above
expression. The expansion coefficients $A_{\kappa }^{\lambda }$ and $A_{\kappa }^{\mu }$ are easily computed by considering the spectral
representations (\ref{spec}),
\begin{eqnarray}
A_{\kappa }^{\lambda }&=\frac{1}{4^{N}}\Tr\{\rho(Z_{\kappa }X_{\lambda \kappa
})^{\dagger }\}=\frac{1}{4^{N}}c_{\kappa ,\lambda }\sum_{\eta
}i^{3\mathrm{T}_{4}(\kappa \eta )}p_{\eta }^{\lambda },\\ \label{alamb}
\tilde{A}_{\kappa }^{\mu }&=\frac{1}{4^{N}}\Tr\{\rho (Z_{\mu \kappa }X_{\kappa})^{\dagger}\}=\frac{1}{4^{N}}c_{\kappa ,\mu }\sum_{\eta }i^{3\mathrm{T}_{4}(\kappa\eta )}\tilde{p}_{\eta}^{\mu},  \label{amu}
\end{eqnarray}
where $p_{\kappa }^{\lambda }$ and $\tilde{p}_{\kappa }^{\mu }$ are the
measured probabilities $p_{\kappa }^{\lambda }=\langle \psi _{\kappa}^{\lambda }|\rho |\psi _{\kappa }^{\lambda }\rangle $, $\tilde{p}_{\kappa}^{\mu }=\langle \tilde{\psi}_{\kappa }^{\mu }|\rho |\tilde{\psi}_{\kappa}^{\mu }\rangle $.

Substituting (\ref{amu}) and (\ref{alamb}) into (\ref{redun}) and taking into
account the summation rules on the ring,
\begin{eqnarray*}
\sum_{\alpha \in GR(4,N)}i^{\mathrm{T}_{4}(\alpha \kappa )}=4^{N}\delta _{\kappa,0},\quad \sum_{\alpha \in (2)}i^{\mathrm{T}_{4}(\alpha \kappa )}=2^{N}\delta _{\bar{\kappa},0},
\end{eqnarray*}
we arrive at the following tomographic expression for the $N$-ququart density matrix,
\begin{eqnarray}
\rho =\sum_{\mu \in (2)}\sum_{\kappa \in GR(4,N)}\tilde{C}_{\kappa }^{\mu }|\tilde{\psi}_{\kappa }^{\mu }\rangle \langle \tilde{\psi}_{\kappa }^{\mu
}|+\sum_{\lambda ,\kappa \in GR(4,N)}C_{\kappa }^{\lambda }|\psi _{\kappa
}^{\lambda }\rangle \langle \psi _{\kappa }^{\lambda }|-\frac{\mathbb{I}}{2^{N}},  \label{tomografiaNquadrits}
\end{eqnarray}
where,
\begin{eqnarray}
\tilde{C}_{\kappa }^{\mu }&=\tilde{p}_{\kappa }^{\mu }-\frac{2^{N}-1}{4^{N}}%
\sum_{\gamma \in (2)}\tilde{p}_{\kappa +\gamma }^{\mu }, \\ \label{coefi}
C_{\kappa }^{\lambda }&=p_{\kappa}^{\lambda }-\frac{2^{N}-1}{4^{N}}\sum_{\gamma \in (2)}p_{\kappa +\gamma }^{\lambda }.  \label{coefi1}
\end{eqnarray}
The reconstruction equation (\ref{tomografiaNquadrits}) is similar to the
corresponding one for the qubit MUB tomography (\ref{qbformula}), except
that the coefficients are linear combinations of the measured probabilities.
The probabilities $p_{\kappa }^{\lambda }$ and $\tilde{p}_{\kappa }^{\mu }$
are not independent and apart from the normalization conditions similar to (\ref{NC}), they satisfy the following relations directly followed from (\ref{RC1}) - (\ref{RC2}),
\begin{eqnarray}
\sum_{\gamma \in (2)}p_{\bar{\kappa}+\gamma }^{\bar{\lambda}+\delta
}=\sum_{\gamma \in (2)}p_{\bar{\kappa}+\gamma }^{\bar{\lambda}},\quad
\sum_{\gamma \in (2)}\tilde{p}_{\bar{\kappa}+\gamma }^{\mu }=\sum_{\gamma
\in (2)}\tilde{p}_{\bar{\kappa}+\gamma }^{0},  \label{redundancia}
\end{eqnarray}
where $\delta ,\mu \in (2)$, explicitly reflecting the overcompleteness of
the present measurement scheme.

\subsection{Error estimation}

Projectors on the elements of the bases (\ref{4zx}) and (\ref{4xz}) can be
considered as output channels corresponding to specific setups, specified by
$\lambda \in GR(4,N)$ and $\mu \in (2)$. Performing a finite number of
measurements, $M$, in each setup one counts the frequencies $f_{\kappa
}^{\lambda }=m_{\kappa }^{\lambda }/M$, ($\tilde{f}_{\kappa }^{\mu }=\tilde{m}_{\kappa }^{\nu }/M$), where $m_{\kappa }^{\lambda }$, ($\tilde{m}_{\kappa}^{\mu }$) is the number of times the state is projected into $|\psi
_{\kappa }^{\lambda }\rangle $, ($|\tilde{\psi}_{\kappa }^{\mu }\rangle $),
obeying the multinomial statistics $\propto \Pi _{\kappa }(p_{\kappa}^{\lambda })^{m_{\kappa }^{\lambda }}$, ($\propto \Pi _{\kappa }(\tilde{p}_{\kappa }^{\mu })^{\tilde{m}_{\kappa }^{\mu }}$).

The accuracy of the reconstruction scheme (\ref{tomografiaNquadrits}) can be
measured by the statistical average of the Hilbert-Schmidt distance between
the real ($\rho $) and estimated ($\rho_{est})$ density matrices \cite{checos},
\begin{eqnarray}
\langle \mathcal{E}^{2}\rangle =\langle \Tr(\rho -\rho_{est})^{2}\rangle .
\label{hs}
\end{eqnarray}
In terms of \textit{independent} probabilities (see \ref{mse}) the square
error (\ref{hs}) is represented as,
\begin{eqnarray*}
\langle \mathcal{E}^{2}\rangle =\Delta \mathbf{p}^{T}Q\Delta \mathbf{p},
\end{eqnarray*}
where $\Delta \mathbf{p}=\mathbf{p}-\mathbf{p}_{est}$ is the vector of the
difference between real and estimated (independent) probabilities and the
matrix $Q$ is expressed in terms of scalar products between the projectors
appearing in (\ref{tomografiaNquadrits}). It is worth noting that for $N$-qubits the $Q$ matrix has a trivial form,
\begin{eqnarray*}
Q=\bigoplus_{\lambda \in \mathbb{F}_{2^{2N}}}Q^{(\lambda )}\quad q_{\kappa
,\eta }^{\lambda }=1+\delta _{\kappa ,\eta },\quad \kappa ,\eta ,\lambda \in
\mathbb{F}_{2^{2N}}.
\end{eqnarray*}
This is not the case for $N$-ququarts, since probabilities labelled with the
same $\bar{\lambda}$ are related according to (\ref{redundancia}) (see the
explicit expressions for the $Q$-matrix in \ref{mse}).

The minimum square error (\ref{hs}) is fixed by the Cram\'{e}r-Rao bound
\cite{helstrom},
\begin{eqnarray}
\langle \mathcal{E}^{2}\rangle \geq \Tr\left( Q\mathcal{F}^{-1}\right),
\label{CR}
\end{eqnarray}
where $\mathcal{F}$ is the Fisher matrix. The algebraic structure of the
bases (\ref{4zx})-(\ref{4xz}) allows to compute explicitly both $Q$ and
Fisher matrices, and thus, analytically estimate the minimum mean square
error, see \ref{mse}.

It is instructive to compare the performance of tomographic protocols for i)
$N$-ququarts (\ref{tomografiaNquadrits}), ii) $2N$ qubits (\ref{qbformula}),
and iii) informationally complete symmetric POVMs (SIC-POVMs) \cite{sic} in
the Hilbert space $\mathcal{H}$ of dimension $d=4^{N}$.

\begin{itemize}
\item For a system of $2N$ qubits the lower bound on the MSE is \cite{checos},
\begin{eqnarray}
\langle \mathcal{E}_{MUB}^{2}\rangle _{\min }=2^{2N}+1-\sum_{\kappa \in
\mathbb{F}_{2^{2N}}}\left( \sum_{\lambda \in \mathbb{F}_{2^{2N}}}(p_{\kappa
}^{\lambda })^{2}+(\tilde{p}_{\kappa }^{0})^{2}\right) ,  \label{errorMUB}
\end{eqnarray}
where $p_{\kappa }^{\lambda }=\langle \psi _{\kappa }^{\lambda }|\rho |\psi
_{\kappa }^{\lambda }\rangle $ and $\tilde{p}_{\kappa }=\langle \tilde{\kappa}|\rho |\tilde{\kappa}\rangle $ are the probabilities of projecting on the unbiased bases (\ref{MUB2Mb}) and (\ref{2mub}).

\item The lower bound of the MSE for SIC-POVMS in dimension $d=4^{N}$ is,
\cite{povm1,englert}
\begin{eqnarray}
\left\langle \mathcal{E}_{SIC}^{2}\right\rangle _{\min
}=4^{2N}+4^{N}-1-\Tr(\rho ^{2}),  \label{errorSIC}
\end{eqnarray}
where $p_{k}$, $k=1,...,4^{N}$, represents the probability of obtaining the
outcome associated with $\Pi _{k}$, which are the first-rank projectors $%
\{\Pi _{k},k=1,..,4^{2N}\}$, such that, $\Tr(\Pi _{k}\Pi _{l})=1/(4^{N}+1)$, $k\neq l$.
\end{itemize}

In table \ref{msetable} we show $\sqrt{\langle \mathcal{E}^{2}\rangle _{\min
}}$ averaged over $10^{3}$ \textbf{pure states, randomly generated using the Fubini-Study metric \textbf{(second column)} and separately over $10^{3}$ mixed states randomly generated using the Hilbert-Schmidt metric \textbf{(third column)}} for i) one and two
ququarts; ii) two and four qubits and iii) $\dim \mathcal{H}=4$ and $16$.
One can observe that the ququart tomography exhibits a better performance
that MUB quibit and SIC POVM reconstructions in the Hilbert spaces of the
same dimension. This can be attributed to the redundancy in the acquired
data needed for the reconstruction protocol (\ref{tomografiaNquadrits}). In
addition, it is worth noting that the amount of non-local gates required for
the generation of $N$-ququart bases (\ref{4zx}) and (\ref{4xz}) is
significantly lower than the one for the generation of $2N$-qubit MUBs (\ref%
{2mub}). For instance, for 2 ququarts one 14 $CNOT_{4}$ gates (\ref{4CNOT})
are needed against 40 $CNOT_{2}$ gates required in the four-qubit case \cite%
{klimov1}, \cite{max}.

\begin{table}
\caption{Minimum square error $\protect\sqrt{\langle \mathcal{E}^{2}\rangle_{\min }}$ averaged over $10^{3}$ pure states, randomly generated using the Fubini-Study metric \textbf{(second column)} and separately over $10^{3}$ mixed states randomly generated using the Hilbert-Schmidt metric \textbf{(third column)} for different
tomographic schemes.}\label{msetable}
\begin{center}
\begin{tabular}{|c|c|c|}
\hline
Tomographic scheme & $\sqrt{\langle \mathcal{E}^{2}\rangle _{\min }}$ (pure) & $\sqrt{\langle \mathcal{E}^{2}\rangle _{\min }}$ (mixed) \\
\hline
\textit{single ququart MUB-like} & 1.72 & 1.84\\ \hline
\textit{MUB 2 qubits} & 1.88 & 1.95\\ \hline
\textit{d=4 SIC-POVM} & 4.24 & 4.44\\ \hline\hline
\textit{2 ququarts MUB-like} & 3.16 & 3.54\\ \hline
\textit{MUB 4 qubits} & 3.87 & 3.98\\ \hline
\textit{\ d=16 SIC-POVM} & 16.43 & 16.49\\ \hline
\end{tabular}
\end{center}
\end{table}

\section{Conclusions}

The standard algebraic approach, based on the finite field structure, to
construct a complete set of mutually unbiased bases can be applied only to
systems of power-prime dimensions \cite{wootters}. Such bases are employed
in the optimal tomographic scheme, characterized by a non-redundant set of
measurements. In the simplest multipartite case of $N$-ququarts such an
approach is not feasible, since the underlying algebraic structure Galois
rings does not allow to construct genuine set of MUBs. However, the
algebraic structure of the bases (\ref{4zx})-(\ref{4xz}) is sufficiently
simple so that an explicit reconstruction protocol can be found. In contrast
to the qubit case, it requires redundant measurements. The extra information
contained in these measurements leads to\ a reduction of the statistical
error in comparison with the qubit MUB and SIC POVM schemes in a Hilbert
space of the same dimension.

\appendix

\section{Finite fields}

\label{fields}

A set $\mathcal{R}$ is a commutative \textit{ring} if two commutative and
associative binary operations: addition ($+$) and multiplication ($\cdot $)
are defined. $(\mathcal{R},+)$ forms a group, i.e. for any $a\in \mathcal{R}$
there exist $-a\in \mathcal{R}$ such that $a+(-a)=0$. The set $\mathbb{Z}%
_{N}=\{0,1,\ldots ,N-1\}$ forms a ring, where all algebraic operations are $%
\mathrm{mod}~N$.

A \textit{field} $\mathbb{F}$ is a commutative ring with division, i.e. for
any $a\in \mathbb{F}$ there exists $a^{-1}\in \mathbb{F}$ so that $%
a^{-1}a=aa^{-1}=I$ (excluding the zero element). Elements of a field form
groups with respect to addition in $\mathbb{F}$ and multiplication in $%
\mathbb{F}^{\ast }=\mathbb{F}-\left\{ 0\right\} $. The set $\mathbb{Z}%
_{2}=\{0,1\}$ is a field, while the set $\mathbb{Z}_{4}=\{0,1,2,3\}$ does
not have a finite field structure since $2$ has not inverse in $\mathbb{Z}%
_{4}$.

A \textit{finite field} is a field with a finite number of elements, its
\textit{characteristic} is the smallest integer $p$, so that,
\begin{eqnarray*}
p\cdot 1=\underbrace{1+1+..+1}_{\mbox{\scriptsize $p$ times}}=0,
\end{eqnarray*}
and it is always a prime number.

Any finite field contains a prime subfield $\mathbb{Z}_{p}$ and has $p^{N}$
elements, where $N$ is a natural number. The finite field containing $p^{N}$
elements is unique and is usually referred as Galois field, $\mathbb{F}%
_{p^{N}}$. $\mathbb{F}_{p^{N}}$ is an extension of degree $N$ of $\mathbb{Z}%
_{p}$, i.e. elements of $\mathbb{F}_{p^{N}}$ can be obtained with $\mathbb{Z}_{p}$ and all the roots of an irreducible (i.e. non-factorable in $\mathbb{Z}_{p}$) polynomial of degree $N$ with coefficients in $\mathbb{Z}_{p}$.

The multiplicative group of $\mathbb{F}_{p^{N}}:\mathbb{F}_{p^{N}}^*=\mathbb{F}_{p^{N}}-\left\{ 0\right\} $ is cyclic $\theta ^{p^{N}}=\theta $, $\theta \in \mathbb{F}_{p^{N}}$. The generators of this group are called
\textit{primitive elements} of the field. A primitive element of $\mathbb{F}_{p^{N}}$ is a root of an irreducible polynomial of degree $N$ over $\mathbb{Z}_{p}$. This polynomial is called a \textit{primitive polynomial}$h(x)$.

The map $\sigma (\alpha )=\alpha^{p}$ on $\mathbb{F}_{p^{N}}$ (over $\mathbb{Z}_{p}$) is a linear automorphism of $\mathbb{F}_{p^{N}}$, called
\textit{Frobenius automorphism}. Frobenius automorphisms leave the prime
subfield $\mathbb{Z}_{p}$ invariant. The \textit{trace operation} $\tr:\mathbb{F}_{p^{N}}\rightarrow \mathbb{Z}_{p}$ is defined as,
\begin{eqnarray*}
\tr(\alpha )=\alpha +\sigma (\alpha )+...+\sigma ^{N-1}(\alpha)=\alpha +\alpha ^{p}+\alpha ^{p^{2}}+...+\alpha ^{p^{N-1}},\quad \alpha
\in \mathbb{F}_{p^{N}}.
\end{eqnarray*}
Any $\alpha \in \mathbb{F}_{p^{N}}$ can be written as,
\begin{eqnarray*}
\alpha =a_{1}\theta _{1}+\cdots +a_{N}\theta _{N},
\end{eqnarray*}
where $a_{i}\in \mathbb{Z}_{p}$ and $\{\theta _{1},...,\theta _{N}\}$ is a
basis for $\mathbb{F}_{p^{N}}$. Two bases $\left\{ \theta _{1},\ldots,\theta _{N}\right\}$ and $\left\{ \theta _{1}^{\ast },\ldots ,\theta_{N}^{\ast }\right\} $ are \textit{dual} if $\tr\left( \theta_{i}\theta _{j}^{\ast }\right) =\delta _{ij}$. A basis which is dual to itself is called \textit{self-dual basis}, $\tr\left(\theta_{i}\theta _{j}\right) =\delta _{ij}$. For finite fields of characteristic $2$, i.e. $\mathbb{F}_{2^{N}}$, there always exists a self-dual basis.

Example: For $\mathbb{F}_{2^{2}}$, the primitive polynomial is $x^{2}+x+1=0$%
, has the roots $\left\{ \xi ,\xi ^{2}\right\} $. The basis $\left\{\xi,\xi ^{2}\right\} $ is self-dual,
\begin{eqnarray}
\tr\left( \xi \xi \right) =1, & \tr\left( \xi\xi^{2}\right) =0, \\
\tr\left( \xi ^{2}\xi \right) =0, & \tr\left(\xi ^{2}\xi^{2}\right) =1.
\end{eqnarray}

\section{Galois Rings}

\label{galois}

We focus on commutative rings with an identity element, which main
difference from a field is the existence of elements without a
multiplicative inverse. An element $a\neq 0\in \mathcal{R}$ is called
\textit{zero divisor} if there exists $b\neq 0\in \mathcal{R}$, such that $ab=0\in \mathcal{R}$. For instance, in $\mathbb{Z}_{4}$ the set of zero
divisors is $\{2\}$, while for $\mathbb{Z}_{8}$ is $\{2,4,6\}$.

An non empty set $\mathcal{I}\subset \mathcal{R}$ is called an \textit{ideal}
of $\mathcal{R}$, if for any $a,b\in \mathcal{I}$ and $r\in \mathcal{R}$,
the sum $a+b\in \mathcal{I}$ and $ra\in \mathcal{I}$. The ideal $Ra=\{ra:r\in \mathbb{R}\}\equiv $ $(a)$, for some $a\in \mathcal{R}$ is called \textit{principal ideal} (generated by $a$). The set $(2)=\{0,2\}\subset\mathbb{Z}_{4}$ is the principal ideal, while in $\mathbb{Z}_{8}$ the principal ideals are the sets $(2)=\{0,2,4,6\}$, and $(4)=\{0,4\}$. The \textit{maximal ideal} is a principal ideal, such that any other principal ideal is proper. The maximal ideal of $\mathbb{Z}_{8}$ is $(2)$. In this paper we will refer indistinctly as zero divisors to the set of zero
divisors and zero, i.e. the maximal ideal $(2)$.

A \textit{Galois ring} is a finite ring with an identity element, such that
the set of its zero divisors added with zero form a maximal principal ideal $(p)$, where $p$ is always a prime number. $\mathbb{Z}_{p^{s}}$ is a Galois
ring with $p^{s}$ elements, for any $p$ prime and $s$ positive integer; its
principal ideals are $(1),(p),(p^{2}),\ldots (p^{s-1})$, and the unique
maximal ideal is $(p)$.

Any element of $\mathbb{Z}_{p^{s}}$ can be written in the form,
\begin{eqnarray*}
c_{1}+c_{2}p+c_{3}p^{2}+\cdots +c_{s}p^{s-1},
\end{eqnarray*}
where $c_{i}\in \mathbb{Z}_{p}$. The \textit{characteristic} of a Galois
ring is the smallest integer such that,
\begin{eqnarray*}
p^{s}\cdot 1=\underbrace{1+1+..+1}_{\mbox{\scriptsize $p^s$ times}}=0,
\end{eqnarray*}
and it is always a power of a prime number.

The \textit{bar map} is a homomorphism, $~\bar{}:\mathbb{Z}_{p^{s}}\rightarrow \mathbb{Z}_{p}=\mathbb{F}_p$, given by,
\begin{eqnarray*}
\overline{a}=a~\mathrm{mod}~ p,\quad a\in \mathbb{Z}_{p^{s}},
\end{eqnarray*}
or equivalently,
\begin{eqnarray*}
\overline{c_{1}+c_{2}p+c_{3}p^{2}+\cdots +c_{s}p^{s-1}}\rightarrow c_{1}.
\end{eqnarray*}
A Galois ring of characteristic $p^{s}$ and the cardinality $p^{sN}$ is
denoted as $GR(p^{s},N)$, $\mathbb{Z}_{p^{s}}\subset G(p^{s},N)$. The bar
map extended to the polynomial ring $\mathbb{Z}_{p^{s}}[x]$ over $\mathbb{Z}%
_{p^{s}}$ gives the polynomial ring $\mathbb{F}_{p}[x]$ over $\mathbb{F}_{p}$:
\begin{eqnarray*}
a_{1}+a_{2}x+\cdots +a_{N}x^{N-1}\rightarrow \overline{a_{1}}+\overline{a_{2}%
}x+\cdots +\overline{a_{n}}x^{N-1}.
\end{eqnarray*}
The Galois ring $GR(p^{s},N)$ is an extension of degree $N$ of $\mathbb{Z}%
_{p^{s}}$, i.e. elements of $GR(p^{s},N)$ are obtained adding to $\mathbb{Z}%
_{p^{s}}$ all the roots of a basic monic irreducible (non-factorizable in $\mathbb{Z}_{p^{s}}$ , whose bar map is monic irreducible in $\mathbb{F}_{p^{N}}$) polynomial $f_{s}(x)$ of degree $N$ that divides $x^{p^{N}}-1$
with coefficients in $\mathbb{Z}_{p^{s}}$. There are $p^{N}$ zero divisors
in $GR(p^{s},N)$.

Any element $\alpha \in GR(p^{s},N)$ can be uniquely written in the \textit{additive representation:}
\begin{eqnarray*}
\alpha =a_{1}+a_{2}\xi +a_{3}\xi ^{2}+\cdots +a_{N}\xi ^{N-1},
\end{eqnarray*}
being $\xi $ a root of $f_{s}(x)$, and $a_{i}\in \mathbb{Z}_{p^{s}}$.

\subsection{The trace map and the self-dual basis}\label{galois1}

The generalized trace map from $GR(p^{s},N)$ to $\mathbb{Z}_{p^{s}}$, which
is an additive operation given by
\begin{eqnarray*}
\mathrm{T}_{p^{s}}(\alpha )=\sum_{i=0}^{N-1}\phi ^{i}(\alpha )=\alpha +\phi (\alpha)+...+\phi ^{N-1}(\alpha ),
\end{eqnarray*}
where $\alpha =a_{0}+a_{1}\xi +...+a_{N-1}\xi ^{N-1}$ and $\phi (\alpha )$
is the generalized Frobenius automorphism,
\begin{eqnarray*}
\phi^{i}(\alpha )=a_{0}+a_{1}\xi ^{p^{i}}+...+a_{N-1}\xi ^{p^{i(N-1)}},
\end{eqnarray*}
where $\phi ^{0}(\alpha )=\alpha$.

As well as in the case of the field $\mathbb{F}_{p^{N}}$, the ring $GR(p^{s},N)$ can be considered as a linear space spanned by a basis $\{\theta _{1},\theta _{2},...,\theta_{N}\}$, which is \textit{self-dual} if $\mathrm{T}_{p^{s}}(\theta _{i}\theta_{j})=\delta _{ij}$. Self-dual bases do not always exist. For instance,there is no self-dual basis in $GR(4,2)$ .

\subsection{2-adic Representation}\label{galois2}

Let us consider the Galois ring $GR(2^{s},N)$ and $\xi$ is a root of the
basic primitive polynomial $f_{s}(x)$ of degree $N$ over $\mathbb{Z}_{2^{s}}$. The subset (the Teichmuller set) $\mathcal{T}_{s}=\{0,\xi ,\xi^{2},...,\xi ^{2^{N}-1}=1\}\subset GR(2^{s},N)$ allows to represent any element $\alpha \in GR(2^{s},N)$ as,
\begin{eqnarray}
\alpha =a_{1}+2a_{2}+4a_{3}+\cdots +2^{s-1}a_{s},\quad a_{i}\in \mathcal{T}_{s},
\end{eqnarray}
which is \textit{2-adic} representation of the ring $GR(2^{s},N)$, and $\overline{a}_i\in\mathbb{F}_{2^N}$, for all $a_i\in\mathcal{T}_s$. In this
representation, invertible elements of the ring (that are called units) are
such that $a_{1}\neq 0$ and the zero divisors correspond to $a_{1}=0$. The
application of the bar map becomes trivial in this representation: $\overline{\alpha }=\overline{a}_{1}$.

\subsection{Hensel Lift}

A basic monic irreducible polynomial $f_{s}(x)$ on $\mathbb{Z}_{2^{s}}[x]$
can be considered as the \textit{Hensel Lift} of a monic irreducible
polynomial $f_{1}(x)=\overline{f_{s}(x)}$ on $\mathbb{Z}_{2}[x]$. Such a
lifting is unique. The lifting can be divided in steps: there is a unique
Hensel lift from $f_{2^{s}}|(x^{2^{N}}-1)$ in $\mathbb{Z}_{2^{s}}[x]$ to $f_{s+1}(x)|(x^{2^{N}}-1)$ in $\mathbb{\ Z}_{2^{s+1}}[x]$, so that $f_{s+1}(x)~\mathrm{mod}~2^{s}=f_{s}(x).$ The roots of $f_{{s}}$ and $f_{s+1}(x)
$ define $GR(2^{s},N)$ and $GR(2^{s+1},N)$ respectively \cite{ecodes1}. The
corresponding trace operations are related as,
\begin{eqnarray}
\mathrm{T}_{2^{s+1}}(\alpha )~\mathrm{mod}~2^{s}=\mathrm{T}_{2^{s}}(\alpha ~\mathrm{mod}~2^{s}),
\label{hens}
\end{eqnarray}
where $\alpha \in GR(2^{s+1},N)$ and $\alpha ~\mathrm{mod}~2^{s}\in GR(2^{s},N)$.

For instance, i) the Hensel lift $Z_{2}[x]\rightarrow Z_{2^{2}}[x]$ of the
irreducible polynomial $x^{2}+x+1$ that divides $x^{4}-1$ is $x^{2}+x+1$, and, ii) the Hensel lift $Z_{2^{2}}[x]\rightarrow Z_{2^{3}}[x]$ of the irreducible polynomial in $x^{2}+x+1$ that divides $x^{4}-1$ is $x^{2}+x+1$.

\subsection{Solution of equation $c_{\protect\alpha +\protect\gamma ,\protect\lambda }c_{\protect\gamma ,\protect\lambda }^*=c_{\protect\alpha ,\protect\lambda }\protect\omega _{s}^{-\mathrm{T}_{2^{s}}(\protect\alpha \protect\gamma \protect\lambda )}$}\label{galois4}

Consider the following equation,
\begin{eqnarray}
c_{\alpha +\gamma ,\lambda }c_{\gamma ,\lambda }^*=c_{\alpha ,\lambda
}\omega _{s}^{-\mathrm{T}_{2^{s}}(\alpha \gamma \lambda )},\quad |c_{\gamma ,\lambda
}|=1,\quad c_{0,\lambda }=1,  \label{generalphase}
\end{eqnarray}
where $\alpha ,\gamma ,\lambda \in GR(2^{s},N)$ and $\omega _{s}$ is $2^{s}$-th root of unity. The functional Eq. (\ref{generalphase}) admits a simple
solution if the subindices of $c_{\gamma ,\lambda }$ are considered as elements of the set,
\begin{eqnarray*}
\mathcal{T}_{s+1}=GR(2^{s+1},N)/(2^{s}),
\end{eqnarray*}
which has the same cardinality as $GR(2^{s},N)$. Therefore, the elements of $\mathcal{T}_{s+1}$ are in one-to-one correspondence with $GR(2^{s},N)$, and,
\begin{eqnarray*}
\mathcal{T}_{s+1}~\mathrm{mod}~2^{s}=GR(2^{s},N).
\end{eqnarray*}
Then, Eq.(\ref{generalphase}) can be rewritten according to (\ref{hens}) as
follows,
\begin{eqnarray*}
c_{\alpha +\gamma ,\lambda }=c_{\alpha ,\lambda }c_{\gamma ,\lambda }\omega
_{s+1}^{-\mathrm{T}_{2^{s+1}}(\alpha \gamma \lambda )},\quad \alpha ,\gamma
,\lambda \in \mathcal{T}_{s+1}\subset GR(2^{s+1},N),
\end{eqnarray*}
such that $\alpha ,\gamma ,\lambda ~\mathrm{mod}~2^{s}\in GR(2^{s},N)$.
Imposing the condition $c_{0,\lambda }=c_{2^{s}\gamma ,\lambda }=1$, we
immediately arrive at the following solution,
\begin{eqnarray}
c_{\gamma ,\lambda }=\omega _{s+1}^{-\mathrm{T}_{2^{s+1}}(\lambda \gamma^{2})}.
\label{cG}
\end{eqnarray}
For instance, in the qubit case, corresponding to $s=1$, $c_{\gamma
,\lambda }=i^{3\mathrm{T}_{4}(\lambda \gamma ^{2})}$ is obtained, where $\gamma,\lambda \in \mathcal{T}_{2}=GR(4,N)/(2)\subset GR(4,N)$ and $\mathcal{T}_{2}~\mathrm{mod}~2=GR(2,N)=\mathbb{F}_{2^{N}}.$

\subsection{GR(4,2)}

The basic monic irreducible polynomial $h(x)=x^{2}+x+1$ in $\mathbb{Z}_{4}[x] $ uniquely defines the elements of $GR(4,2)$ as the residue class $\mathbb{Z}_{4}[x]/(h(\xi ))$, where $\xi $ is a root of $h(x)$. In the $2$-adic representation, the $4^{2}$ elements are split into two sets,

\begin{enumerate}
\item Units: $\alpha=1+2b, \xi+2b, \xi^2+2b$

\item Zero divisors: $\alpha=2b$
\end{enumerate}
where $b\in \mathcal{T}_{2}=\{0,1=\xi ^{3},\xi ,\xi ^{2}\}$. A suitable
representation of $GR(4,2)$ as elements of $\mathbb{Z}_{4}\times \mathbb{Z}_{4}$ is $\kappa =k_{0}\xi +k_{1}\xi ^{2}$, where $k_{0},k_{1}\in GR(4,N)$
and $\{\xi ,\xi ^{2}\}$ form a (non self-dual) basis, i.e.,
\begin{eqnarray*}
T_{4}(\xi ^{i}\xi ^{j})=\delta _{i,j}+2,
\end{eqnarray*}
where,
\begin{eqnarray*}
\mathrm{T}_{4}(\xi ^{i}\xi ^{j}\xi ^{k})=\left\{
\begin{array}{ll}
2 & \mbox{if $i = j = k$}; \\
3 & \mbox{otherwise}.%
\end{array}%
\right.
\end{eqnarray*}
In table \ref{bases2qq} the expansion in the basis $\{\xi ,\xi ^{2}\}$ and $2
$-adic representation, for the irreducible polynomial $\xi ^{2}+\xi +1=0$
are present.
\begin{table}
\caption{Expansion of elements of $GR(4,2)$ in the basis $\{\protect\xi ,%
\protect\xi ^{2}\}$ and the corresponding 2-adic representation. }
\label{bases2qq}
\begin{center}
\begin{tabular}{|c|c|c|c|}
\hline
2-adic & Basis expansion & 2-adic & Basis expansion \\ \hline
$0$ & $0$ & $\xi $ & $\xi $ \\ \hline
$2$ & $2\xi +2\xi ^{2}$ & $\xi +2$ & $3\xi +2\xi ^{2}$ \\ \hline
$2\xi $ & $2\xi $ & $3\xi $ & $3\xi $ \\ \hline
$2\xi ^{2}$ & $2\xi ^{2}$ & $\xi +2\xi ^{2}$ & $\xi +2\xi ^{2}$ \\ \hline
$1$ & $3\xi +3\xi ^{2}$ & $\xi ^{2}$ & $\xi ^{2}$ \\ \hline
$3$ & $\xi +\xi ^{2}$ & $\xi ^{2}+2$ & $2\xi +3\xi ^{2}$ \\ \hline
$1+2\xi $ & $\xi +3\xi ^{2}$ & $\xi ^{2}+2\xi $ & $\xi ^{2}+2\xi $ \\ \hline
$1+2\xi ^{2}$ & $\xi ^{2}+3\xi $ & $3\xi ^{2}$ & $3\xi ^{2}$ \\ \hline
\end{tabular}
\end{center}
\end{table}

\subsection{GR(4,3)}

The ring $GR(4,3)$ is uniquely defined by a root $\xi $ of the {monic basic
irreducible polynomial $h(x)=x^{3}+3x^{2}+2x+3$ over $Z_{4}[x]$. In the $2$%
-adic representation the elements of $GR(4,3)$ have the form $a_{0}+2a_{1}$,
where,
\begin{eqnarray*}
a_{0},a_{1}\in \mathcal{T}_2=\{0,1=\xi ^{7},\xi ,\xi ^{2},\xi ^{3},\xi^{4},\xi ^{5},\xi ^{6}\}.
\end{eqnarray*}
There is a self-dual basis in $GR(4,3)$,
\begin{eqnarray*}
\{\theta _{1}=\xi +2\xi ^{2},\theta _{2}=\xi ^{2}+2\xi ^{4},\theta _{3}=\xi
^{4}+2\xi \}.  \label{baseautodual3}
\end{eqnarray*}

\section{MU-like bases for one ququart}\label{apenC}
Here, six MU-like bases for a single ququart are explicitly constructed.

i) Eigenstates of the set $\{Z^{k},k=0,1,2,3\}$ form the computational basis
$\{|\psi _{k}^{0}\rangle =|k\rangle$, $k=0,1,2,3\}$.

ii) The eigenstates of the set $\{Z^{k}X^{k},k=0,1,2,3\}$ are%
\begin{eqnarray}
| \psi _{0}^{1}|\rangle  &=\frac{1}{2}\left( \frac{1+i}{\sqrt{2}}| 0\rangle +| 1\rangle -\frac{1+i}{\sqrt{2}}| 2\rangle +| 3\rangle \right),\nonumber \\
|\psi _{1}^{1}\rangle  &=\frac{1}{2}\left(|0\rangle +\frac{1+i}{\sqrt{2}}|1\rangle +|2\rangle -\frac{1+i}{\sqrt{2}}|3\rangle \right),\nonumber \\
|\psi _{2}^{1}\rangle  &=\frac{1}{2}\left( -\frac{1+i}{\sqrt{2}}|0\rangle +|1\rangle +\frac{1+i}{\sqrt{2}}|2\rangle +|3\rangle \right),\nonumber \\
|\psi _{3}^{1}\rangle  &=\frac{1}{2}\left(|0\rangle -\frac{1+i}{\sqrt{2}}|1\rangle +|2\rangle +\frac{1+i}{\sqrt{2}}|3\rangle \right) ;\nonumber
\end{eqnarray}

iii) The eigenstates of the set $\{Z^{k}X^{2k},k=0,1,2,3\}$ are%
\begin{eqnarray}
\left\vert \psi _{0}^{2}\right\rangle  &=&\frac{1}{2}\left( (1+i)\left\vert
0\right\rangle +(1-i)\left\vert 2\right\rangle \right) ,~~\left\vert \psi
_{1}^{2}\right\rangle =\frac{1}{2}\left( (1+i)\left\vert 1\right\rangle
+(1-i)\left\vert 3\right\rangle \right) ,  \nonumber \\
\left\vert \psi _{2}^{2}\right\rangle  &=&\frac{1}{2}\left( (1-i)\left\vert
0\right\rangle +(1+i)\left\vert 2\right\rangle \right) ,~~\left\vert \psi
_{3}^{2}\right\rangle =\frac{1}{2}\left( (1-i)\left\vert 1\right\rangle
+(1+i)\left\vert 3\right\rangle \right) ;  \nonumber
\end{eqnarray}

iv) The eigenstates of the set $\{Z^{k}X^{3k},k=0,1,2,3\}$ are%
\begin{eqnarray}
\left\vert \psi _{0}^{3}\right\rangle  &=&\frac{1}{2}\left( -\frac{1-i}{%
\sqrt{2}}\left\vert 0\right\rangle +\left\vert 1\right\rangle +\frac{1-i}{%
\sqrt{2}}\left\vert 2\right\rangle +\left\vert 3\right\rangle \right) ,
\nonumber \\
\left\vert \psi _{1}^{3}\right\rangle  &=&\frac{1}{2}\left( \left\vert
0\right\rangle -\frac{1-i}{\sqrt{2}}\left\vert 1\right\rangle +\left\vert
2\right\rangle +\frac{1-i}{\sqrt{2}}\left\vert 3\right\rangle \right) ,
\nonumber \\
\left\vert \psi _{2}^{3}\right\rangle  &=&\frac{1}{2}\left( \frac{1-i}{\sqrt{%
2}}\left\vert 0\right\rangle +\left\vert 1\right\rangle -\frac{1-i}{\sqrt{2}}%
\left\vert 2\right\rangle +\left\vert 3\right\rangle \right) ,  \nonumber \\
\left\vert \psi _{3}^{3}\right\rangle  &=&\frac{1}{2}\left( \left\vert
0\right\rangle +\frac{1-i}{\sqrt{2}}\left\vert 1\right\rangle +\left\vert
2\right\rangle -\frac{1-i}{\sqrt{2}}\left\vert 3\right\rangle \right) ;
\nonumber
\end{eqnarray}

v) Eigenstates of the set $\{Z^{2k}X^{k},k=0,1,2,3\}$ are
\begin{eqnarray}
\left\vert \tilde{\psi}_{0}^{0}\right\rangle &=&\frac{1}{2}\left( \left\vert
0\right\rangle +\left\vert 1\right\rangle +\left\vert 2\right\rangle
+\left\vert 3\right\rangle \right) ,  \nonumber \\
\left\vert \tilde{\psi}_{1}^{0}\right\rangle &=&\frac{1}{2}\left( \left\vert
0\right\rangle +i\left\vert 1\right\rangle -\left\vert 2\right\rangle
-i\left\vert 3\right\rangle \right) ,  \nonumber \\
\left\vert \tilde{\psi}_{2}^{0}\right\rangle &=&\frac{1}{2}\left( \left\vert
0\right\rangle -\left\vert 1\right\rangle +\left\vert 2\right\rangle
-\left\vert 3\right\rangle \right) ,  \nonumber \\
\left\vert \tilde{\psi}_{3}^{0}\right\rangle &=&\frac{1}{2}\left( \left\vert
0\right\rangle -i\left\vert 1\right\rangle -\left\vert 2\right\rangle
+i\left\vert 3\right\rangle \right) ;  \nonumber
\end{eqnarray}

vi) The eigenstates of the set $\{X^{k},k=0,1,2,3\}$ are

\begin{eqnarray}
\left\vert \tilde{\psi}_{0}^{2}\right\rangle &=&\frac{1}{2}\left( \left\vert
0\right\rangle +i\left\vert 1\right\rangle +\left\vert 2\right\rangle
+i\left\vert 3\right\rangle \right) ,  \nonumber \\
\left\vert \tilde{\psi}_{1}^{2}\right\rangle &=&\frac{1}{2}\left( \left\vert
0\right\rangle -\left\vert 1\right\rangle -\left\vert 2\right\rangle
+\left\vert 3\right\rangle \right) ,  \nonumber \\
\left\vert \tilde{\psi}_{2}^{2}\right\rangle &=&\frac{1}{2}\left( \left\vert
0\right\rangle -i\left\vert 1\right\rangle +\left\vert 2\right\rangle
-i\left\vert 3\right\rangle \right) ,  \nonumber \\
\left\vert \tilde{\psi}_{3}^{2}\right\rangle &=&\frac{1}{2}\left( \left\vert
0\right\rangle +\left\vert 1\right\rangle -\left\vert 2\right\rangle
-\left\vert 3\right\rangle \right) .  \nonumber
\end{eqnarray}

The bases i) and iii), ii) and iv), v) and vi) are not mutually unbiased by
pairs.

\section{MSE for N ququarts}\label{mse}

In this Appendix, an analytical expression for the Cramer-Rao lower bound
\cite{helstrom},
\begin{eqnarray}
\left\langle \mathcal{E}^{2}\right\rangle _{\min }=\Tr(Q\mathcal{F}^{-1}),
\end{eqnarray}
of the mean square error (MSE) proper to the reconstruction protocol (\ref{tomografiaNquadrits}) is obtained.

In order to find the $Q$-matrix Eq. (\ref{tomografiaNquadrits}) is
substituted into (\ref{hs}) and only the independent probabilities are kept
by:

i) removing the $2^{N}+1$ dependent probabilities $\{p_{0}^{\bar{\lambda}},\forall \bar{\lambda}\in \mathcal{T}_{2};\tilde{p}_{0}^{0}\}$ as a
consequence of the normalization conditions,
\begin{eqnarray}
\Delta p_{0}^{\bar{\lambda}}=-\sum_{\kappa \in GR(4,N)/\{0\}}\Delta
p_{\kappa }^{\bar{\lambda}},\quad \Delta \tilde{p}_{0}^{0}=-\sum_{\kappa \in
GR(4,N)/\{0\}}\Delta \tilde{p}_{\kappa }^{0};  \label{prob}
\end{eqnarray}

ii) removing the $4^{N}(2^{N}-1)$ probabilities $\{p_{\bar{\kappa}}^{\bar{\lambda}+\delta },\forall \bar{\lambda},\bar{\kappa}\in \mathcal{T}_{2}\wedge \forall \delta \in (2),\delta \neq 0\}$ due to relation (\ref{redundancia})
\begin{eqnarray}
\Delta p_{\bar{\kappa}}^{\bar{\lambda}+\delta }=-\sum_{\gamma \in
(2)/\{0\}}\Delta p_{\bar{\kappa}+\gamma }^{\bar{\lambda}+\delta
}+\sum_{\gamma \in (2)}\Delta p_{\bar{\kappa}+\gamma }^{\bar{\lambda}},\quad
\delta \in (2),\delta \neq 0;  \label{cond1}
\end{eqnarray}

iii) removing the $2^{N}(2^{N}-1)$ probabilities $\{\tilde{p}_{\bar{\kappa}%
}^{\mu },\forall \bar{\kappa}\in \mathcal{T}_{2}\wedge \forall \mu \in
(2),\mu \neq 0\}$ due to (\ref{redundancia}),
\begin{eqnarray}
\Delta \tilde{p}_{\bar{\kappa}}^{\mu }=-\sum_{\gamma \in (2)/\{0\}}\Delta
\tilde{p}_{\bar{\kappa}+\gamma }^{\mu }+\sum_{\gamma \in (2)}\Delta \tilde{p}_{\bar{\kappa}+\gamma }^{0},\quad \mu \in (2),\mu \neq 0.  \label{cond2}
\end{eqnarray}
Then, the MSE takes the form,
\begin{eqnarray}
\langle \mathcal{E}^{2}\rangle &=\sum_{\bar{\lambda}\in
\mathcal{T}_{4}}\sum_{\kappa,\eta \in GR(4,N)/\{0\}}(Q_{\bar{\lambda}})_{\kappa,\eta}^{\bar{\lambda},\bar{\lambda}}\langle \Delta p_{\kappa }^{\bar{\lambda}}\Delta p_{\eta }^{\bar{\lambda}}\rangle  \nonumber \\
&+\sum_{\bar{\lambda},\bar{\eta}\in \mathcal{T}_{2}}\sum_{\kappa \in
GR(4,N)/\{0\}}\sum_{\delta,\gamma \in (2)/\{0\}}(Q_{\bar{\lambda}})_{\kappa,\bar{\eta}+\gamma}^{\bar{\lambda},\bar{\lambda}+\delta}\langle \Delta
p_{\kappa}^{\bar{\lambda}}\Delta p_{\bar{\eta}+\gamma}^{\bar{\lambda}+\delta}\rangle   \nonumber \\
&+\sum_{\bar{\lambda},\bar{\kappa},\bar{\eta}\in \mathcal{T}_{4}}\sum_{\delta,\delta'\in (2)/\{0\}}\sum_{\gamma,\gamma'\in (2)/\{0\}}(Q_{\bar{\lambda}})_{\bar{\kappa}+\gamma ,\bar{\eta}+\gamma'}^{\bar{\lambda}+\delta ,\bar{\lambda}+\delta'}\langle \Delta p_{\bar{\kappa}+\gamma}^{\bar{\lambda}+\delta}\Delta p_{\bar{\eta}+\gamma'}^{\bar{\lambda}+\delta'}\rangle \nonumber \\
&+\sum_{\kappa,\eta \in GR(4,N)/\{0\}}(\tilde{Q}_{0})_{\kappa,\eta}^{0,0}\langle \Delta \tilde{p}_{\kappa}^{0}\Delta \tilde{p}_{\eta}^{0}\rangle  \nonumber \\
&+\sum_{\kappa \in GR(4,N)/\{0\}}\sum_{\bar{\eta}\in \mathcal{T}_{2}/\{0\}}\sum_{\gamma \in (2)/\{0\}}(\tilde{Q}_{0})_{\kappa ,\bar{\eta}+\gamma }^{0,\mu }\langle \Delta \tilde{p}_{\kappa}^{0}\Delta \tilde{p}_{\bar{\eta}+\gamma}^{\mu}\rangle   \nonumber \\
&+\sum_{\mu,\nu \in (2)/\{0\}}\sum_{\bar{\kappa},\bar{\eta}\in \mathcal{T}_{4}}\sum_{\gamma,\gamma '\in (2)/\{0\}}(\tilde{Q}_{0})_{\bar{\kappa}+\gamma ,\bar{\eta}+\gamma '}^{\mu ,\nu }\langle \Delta \tilde{p}_{\bar{\kappa}+\gamma }^{\mu }\Delta  \tilde{p}_{\bar{\eta}+\gamma'}^{\nu}\rangle, \label{MSEcuartits}
\end{eqnarray}
and the matrix $Q$ has a block diagonal structure,
\begin{eqnarray*}
Q=\bigoplus_{\bar{\lambda}}Q_{\bar{\lambda}}\bigoplus \tilde{Q}_{0}.
\end{eqnarray*}
The non-zero matrix elements of the blocks labelled by $\bar{\lambda}\in\mathcal{T}_{2}$ are,
\begin{eqnarray}
(Q_{\bar{\lambda}})_{\bar{\kappa}+\gamma ,\bar{\eta}+\gamma ^{\prime }}^{\bar{\lambda},\bar{\lambda}} &=\frac{1}{2^{N}}\left[ (4^{N}-2^{N}+1)(\delta_{\kappa ,\eta }+1)\right.   \nonumber \\ 
&+\left. (2^{N}-1)\left[ \delta _{\bar{\kappa},\bar{\eta}}(1-\delta_{\gamma,\gamma ^{\prime }})-2\delta _{\bar{\eta},0}(\delta _{\bar{\kappa},0}+1)\right] \right]   \label{q1} \\
(Q_{\bar{\lambda}})_{\kappa ,\eta }^{\bar{\lambda}+\delta ,\bar{\lambda}+\delta } &=(1-\delta _{\delta,0})(\delta_{\kappa ,\eta }+\delta _{\bar{\kappa},\bar{\eta}}),  \label{q2} \\
(Q_{\bar{\lambda}})_{\kappa ,\eta }^{\bar{\lambda},\bar{\lambda}+\delta }
&=-2(1-\delta _{\bar{\kappa},0}-\delta _{\bar{\eta},0}),  \label{q3}
\end{eqnarray}
where,
\begin{eqnarray*}
\delta _{\bar{\kappa},\bar{\eta}}=\sum_{\gamma \in (2)}\delta _{\kappa ,\eta
+\gamma }.
\end{eqnarray*}
The non-zero elements of the block $\tilde{Q}_{0}$ have the same structure
as (\ref{q1})-(\ref{q3}), with $\overline{\mu }=\overline{\lambda }=0$.

The $[(4^{N})^{2}-1]\times [(4^{N})^{2}-1]$ dimensional Fisher matrix
is formed by $2^{N}+1$ diagonal blocks of dimension $[(4^{N}-1)+2^{N}(2^{N}-1)^{2}]\times [(4^{N}-1)+2^{N}(2^{N}-1)^{2}]$,
\begin{eqnarray}
\mathcal{F}=\bigoplus_{\bar{\lambda}}\mathcal{F}_{\bar{\lambda}}\bigoplus
\tilde{\mathcal{F}}_{0},  \label{direct}
\end{eqnarray}
where each value of $\bar{\lambda}$ defines one diagonal block whose elements are,
\begin{eqnarray}
(\mathcal{F}_{\bar{\lambda}})_{\kappa ,\eta }^{\lambda ,\sigma } &=\frac{1}{M}\left\langle \frac{\partial log\mathcal{L}(\mathbf{n}_{\bar{\lambda}}|\mathbf{p}_{\bar{\lambda}})}{\partial p_{\kappa }^{\lambda }}\frac{\partial
log\mathcal{L}(\mathbf{n}_{\bar{\lambda}}|\mathbf{p}_{\bar{\lambda}})}{\partial p_{\eta }^{\sigma }}\right\rangle ,  \nonumber \\
(\tilde{\mathcal{F}}_{0})_{\kappa ,\eta }^{\mu ,\nu } &=\frac{1}{M}\left\langle \frac{\partial log\mathcal{L}(\tilde{\mathbf{n}}_{0}|\tilde{\mathbf{p}}_{0})}{\partial \tilde{p}_{\kappa }^{\mu }}\frac{\partial log\mathcal{L}(\tilde{\mathbf{n}}_{0}|\tilde{\mathbf{p}}_{0})}{\partial\tilde{p}_{\eta }^{\nu }}\right\rangle ,  \label{Fisherelements}
\end{eqnarray}
also,
\begin{eqnarray}
\mathcal{L}(\mathbf{n}_{\bar{\lambda}}|\mathbf{p}_{\bar{\lambda}})\propto
\Pi _{\kappa }(p_{\kappa }^{\lambda })^{n_{\kappa }^{\lambda }},\quad
\mathcal{L}(\tilde{\mathbf{n}}_{0}|\tilde{\mathbf{p}}_{0})\propto \Pi
_{\kappa }(\tilde{p}_{\kappa }^{\mu })^{\tilde{n}_{\kappa }^{\mu }},
\label{likelihood}
\end{eqnarray}
is the likelihood. The non-zero matrix elements for any of the $2^{N}$
blocks labelled by $\bar{\lambda}\in \mathcal{T}_{2}$ are,
\begin{eqnarray*}
(\mathcal{F}_{\bar{\lambda}})_{\kappa ,\eta }^{\bar{\lambda},\bar{\lambda}}&=\frac{\delta _{\kappa ,\eta }}{p_{\kappa }^{\bar{\lambda}}}+\frac{1}{p_{0}^{\bar{\lambda}}}+(1-\delta _{\bar{\kappa},0})(1-\delta _{\bar{\eta}%
,0})\sum_{\delta ^{\ast }\in (2)}\left( \frac{\delta _{\bar{\kappa}\bar{\eta}}}{p_{\bar{\kappa}}^{\bar{\lambda}+\delta }}+\frac{1}{p_{0}^{\bar{\lambda}+\delta }}\right) ,\quad \kappa ,\eta \neq 0,\\
(\mathcal{F}_{\bar{\lambda}})_{\bar{\kappa}+\gamma ,\bar{\eta}+\gamma
^{\prime }}^{\bar{\lambda}+\delta ,\bar{\lambda}+\delta }&=\frac{\delta
_{\gamma ,\gamma ^{\prime }}}{p_{\bar{\kappa}+\gamma }^{\bar{\lambda}+\delta
}}+\frac{1}{p_{\bar{\kappa}}^{\bar{\lambda}+\delta }},\quad (\mathcal{F}_{\bar{\lambda}})_{\bar{\kappa}+\gamma ,\bar{\kappa}+\gamma ^{\prime }}^{\bar{\lambda},\bar{\lambda}+\delta }=-(1-\delta _{\bar{\kappa},0})\frac{1}{p_{\bar{\kappa}}^{\bar{\lambda}+\delta }},\quad \delta ,\gamma ,\gamma^{\prime}\neq 0;
\end{eqnarray*}
the non-zero elements of the single matrix block labelled by $\bar{\mu}=0$,
are,
\begin{eqnarray*}
(\tilde{\mathcal{F}}_{0})_{\kappa ,\eta }^{0,0}&=\frac{\delta _{\kappa ,\eta }}{\tilde{p}_{\kappa }^{0}}+\frac{1}{\tilde{p}_{0}^{0}}+(1-\delta _{\bar{\kappa},0})(1-\delta _{\bar{\eta},0})\sum_{\nu ^{\ast }\in (2)}\left( \frac{\delta _{\bar{\kappa}\bar{\eta}}}{\tilde{p}_{\bar{\kappa}}^{\nu }}+\frac{1}{\tilde{p}_{0}^{\nu }}\right) ,\quad \kappa ,\eta \neq 0,\\
(\tilde{\mathcal{F}}_{0})_{\bar{\kappa}+\gamma ,\bar{\eta}+\gamma ^{\prime}}^{\mu ,\mu }&=\frac{\delta _{\gamma ,\gamma ^{\prime }}}{\tilde{p}_{\bar{\kappa}+\gamma }^{\mu }}+\frac{1}{\tilde{p}_{\bar{\kappa}}^{\mu }},\qquad (\mathcal{F}_{\mu })_{\bar{\kappa}+\gamma ,\bar{\kappa}+\gamma^{\prime
}}^{0,\mu }=-(1-\delta _{\bar{\kappa},0})\frac{1}{\tilde{p}_{\bar{\kappa}}^{\mu }}\quad \mu ,\gamma ,\gamma ^{\prime }\neq 0.
\end{eqnarray*}
Therefore, the Cr\'{a}mer-Rao lower bound can be written as,
\begin{eqnarray*}
\left\langle \mathcal{E}^{2}\right\rangle _{\min }=\Tr(\tilde{Q}_{0}\tilde{\mathcal{F}}_{0}^{-1})+\sum_{\bar{\lambda}\in \mathcal{T}_{2}}\Tr(Q_{\bar{\lambda}}\mathcal{F}_{\bar{\lambda}}^{-1}).
\end{eqnarray*}

\end{document}